# Modulus and confinement effects on self-repeating, power-amplified snapping of soft, swollen beams

Nolan A. Miller,[1] Laura C. Bradley,[1] Alfred J. Crosby[1]

1. Department of Polymer Science and Engineering. University of Massachusetts Amherst, 120 Governors Drive, Amherst, MA 01003, USA

**Abstract**
Latch-mediated Spring Actuation (LaMSA) is a mechanism found in nature, employed by organisms that generate the highest levels of power density through repeatable, rapid energy release. While LaMSA has been used in engineered systems like archery bows, catapults, and jumping robots, most such technologies require external power for self-repeating motion. Recent advances in soft actuators have demonstrated that engineered gels swollen with a volatile solvent are capable of self-repeating, high specific-power generation by taking advantage of balances between environmental interaction (evaporation) and elasticity. These systems rely upon snap-through instabilities. Due to the complex coupling between material properties and geometry, both of which evolve as self-repeating motion continues, an understanding of how polymer properties and boundary conditions control the lifetime, count, and magnitude of power generating events for a given amount of solvent remains unrealized. We overcome the challenges in characterizing the performance of evaporation-driven, power-generating gels by measuring accumulating force response from evaporation in parallel with the profile deformation of the structure. By optimizing the balance between swelling properties and elasticity, the lifetime of snapping is increased by 445% from previous literature, snapping at a maximum power density of about 87 W/kg. This power is achieved with swollen beams 50 mm in length after 53 mg of solvent had evaporated and is comparable to the power output of adult jumping mantises at 68 W/kg at a similar size scale.[1] We develop scaling relationships that balance Flory-Rehner swelling theory with buckling mechanics to generate insight into optimizing lifetime and power of autonomous power-generating systems.

**Significance Statement**
Autonomous, self-repeating, high-powered actuators are useful for their ability to react to their environment with a unique, powerful mechanical response. We develop design principles that connect the high-powered response of autonomous, evaporation-driven snapping beams to material properties and the spatial confinement of the system. This insight connects buckling mechanics to Latch-Mediated Spring Actuation (LaMSA) principles to better understand the capabilities and limitations in designing actuators driven by their environment.

**Introduction**

Many organisms capable of high power-dense movements, such as trap-jaw ants, mantis shrimp, and Chinese witch-hazel, exhibit diverse mechanisms for powerful actuation.[2–6] The enabling mechanisms of such organisms are unified by a framework, referred to as latch-mediated spring actuation (LaMSA).[7] These systems generate powerful motions by storing energy over time periods that are orders of magnitude greater than the time used to release the energy, enabling powerful actions like mandible closure, striking, or seed dispersal. LaMSA-based plant movements, such as the rapid flytrap leaflet closure, exemplify how the interplay between swelling (actuator), curvature (spring), and buckling (virtual latch) enables repeatable, stimuli-responsive, power-amplified actuations.[8–12] These biological examples have inspired the design of synthetic LaMSA systems capable of high-energy, dynamic behaviors for applications in areas such as capture and launch mechanisms,[13] environmental sensors,[14] and microswimmers.[15] However, only a limited number of synthetic systems can autonomously self-repeat, as natural LaMSA systems do.

Autonomous, self-repeating power-dense movements have been achieved through dynamic, nonlinear processes such as swelling or deswelling of an elastic structure to induce snap-through buckling (a "snap"). In evaporation-induced, power-generating swollen gels, snapping is a result of asymmetric strains at the exposed surfaces of the material. The convex surfaces of buckled shells and beams are strained under tension, driving more solvent to evaporate than at the concave, compressed counterparts. This asymmetry generates an accumulating internal force as solvent evaporates (Figure 1a), driving the system towards an antisymmetric profile. The immediate deformation of the beam is restricted by the stiffness of its buckled geometry at a given solvent concentration (Figure 1b). This resistance establishes a bifurcation point where the asymmetric rate of solvent loss at each surface must overcome the time-dependent stiffness of the structure to snap. Upon snapping, the antisymmetric profile reverses the relative solvent loss rates for the evaporating surfaces, inverting the direction of the evaporative driving force. Autonomous snapping gels can snap numerous times due to this intrinsic reversal process. The power generated by each snap is dependent on the stiffness of the gel at the time of snapping and the solvent concentration gradient. The complex relationship between solvent transport, structure, and deformation that drives snapping performance of these engineered gels is dependent upon their material and geometric properties, yet no study has systematically explored this relationship.

Here we focus on the interplay of elasticity, solvent migration, and geometry in generating the energy required to overcome the snap barrier. We develop scaling relationships, based on Flory-Rehner theory, to relate polymer network properties in a non-swollen state to changes in buckled geometry due to differences in solvent swelling as multiple snap events occur. The modulus of the gel and the structure's initial confinement are used to characterize the optimal conditions for achievable repeatability and power output in autonomously snapping beams driven by solvent evaporation. We also discovered a limit to snap power that is dependent upon the maximum stiffness of the structure. Furthermore, these results uncover new questions about how snap repeatability can be extended, methods to capture the generated energy, and how snapping power can be engineered to mimic self-repeating, power-dense motions seen in nature.

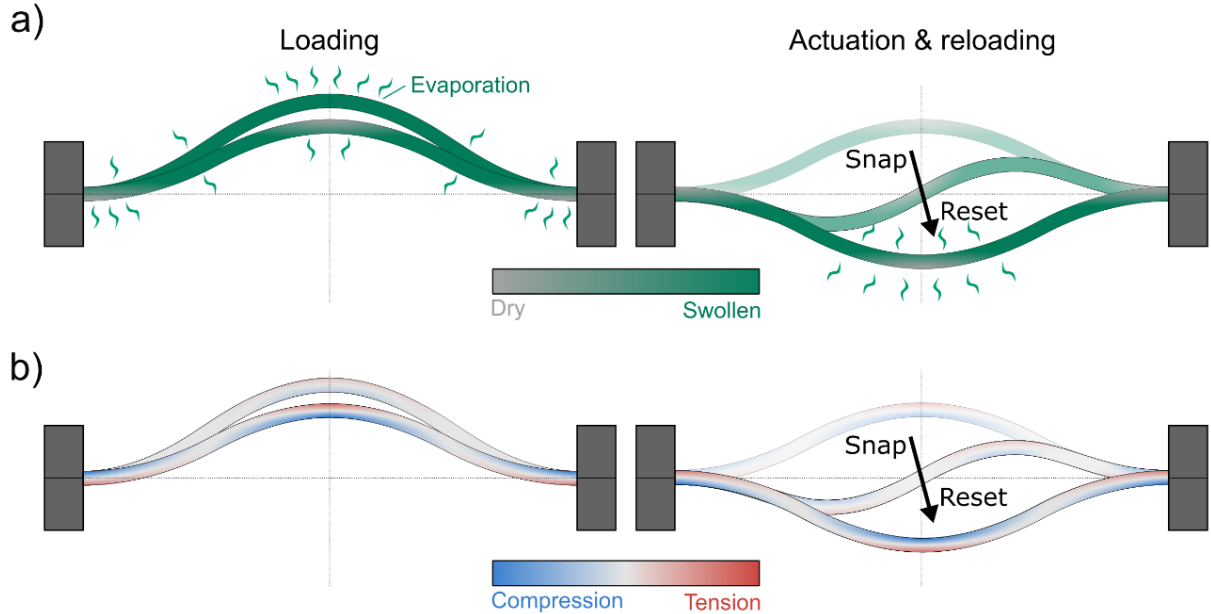

**Figure 1.** Illustration of the combined effects of (a) swelling and (b) buckling strains for the loading phase (left) of a fully swollen, buckled beam as solvent evaporates asymmetrically and the snap-through actuation and reloading (right) of the beam as the evaporative driving force reverses in the antisymmetric profile.

## Results
### Engineering an autonomous, repeatable snapping beam

A flexible, thin beam was made from a crosslinked network of poly(dimethylsiloxane) (PDMS) with four different mixing ratios of a PDMS mixing kit (Sylgard 184). The beams were swollen in toluene to equilibrium, and the fully swollen beam was clamped at fixed ends with 50 mm between the grips and laterally confined (Figure 2a). The beam geometry is described in a Cartesian coordinate system. The x-axis lies laterally along the fixed grips, and the y-axis is oriented vertically to quantify out-of-plane deformation from the x-axis. A single snap event is defined as the moment when the absolute peak displacement of the beam profile changes sign, corresponding to the transition of the peak from one side of the x-axis to the other. Crosslinker concentration was changed to produce networks with different elastic moduli and swelling properties (Figure 2b). The elastic modulus ($E(t)$), lateral confinement ($d(t)$), length ($L(t)$), width ($w(t)$), and thickness ($h(t)$), have transient properties as solvent evaporates from the beam. We denote solvent-concentration dependent variables as a function of time such that at $t = 0$ we consider the buckled beam to be fully swollen and at long times, $t = \infty$, to be fully dried ($h(\infty) = 0.5\ mm, w(\infty) = 5\ mm, L(\infty) = 60\ mm$). The equilibrium swelling ratio ($\lambda$) is measured by the freestanding change in swelling length of the beam, $\lambda = \frac{L(0)}{L(\infty)}$. During the drying process, the profile of the beam is recorded by two cameras: one camera records the entirety of the drying process at 30 fps while a high-speed camera captures individual snaps at 960 fps using motion capture, allowing a displacement threshold to be set to trigger the high-speed camera recording. The highspeed camera allows for the profile of the beam to be analyzed immediately before and after a snap (Figure 2c and Figure 2d). We determine a singular snap to occur when the maximum y-offset along the beam profile changes sign.

The initial confinement of the swollen beam ($d(0)$) is calculated from the change in lateral separation distance between the fixed grips from an initial, swollen length ($L(0) = 50\ mm$) to a confined length ($L_c$) normalized by the initial length of the beam (Equation 1a).

$$d(0) = \frac{L(0) - L_c}{L(0)} \qquad \text{Equation 1a}$$

The confinement of the beam, relative to its stretchability, determines the symmetry for the mode of deformation when an indentation force is applied normal to the center of the beam.[16] We assume the evaporation and diffusion profile to be isotropic such that, for a long slender beam, the contribution of solvent diffusion from material clamped by the grips is negligible. The seven initial confinements tested produce unique profiles giving rise to complex evaporation and concentration gradients throughout the drying process (Figure 2e).

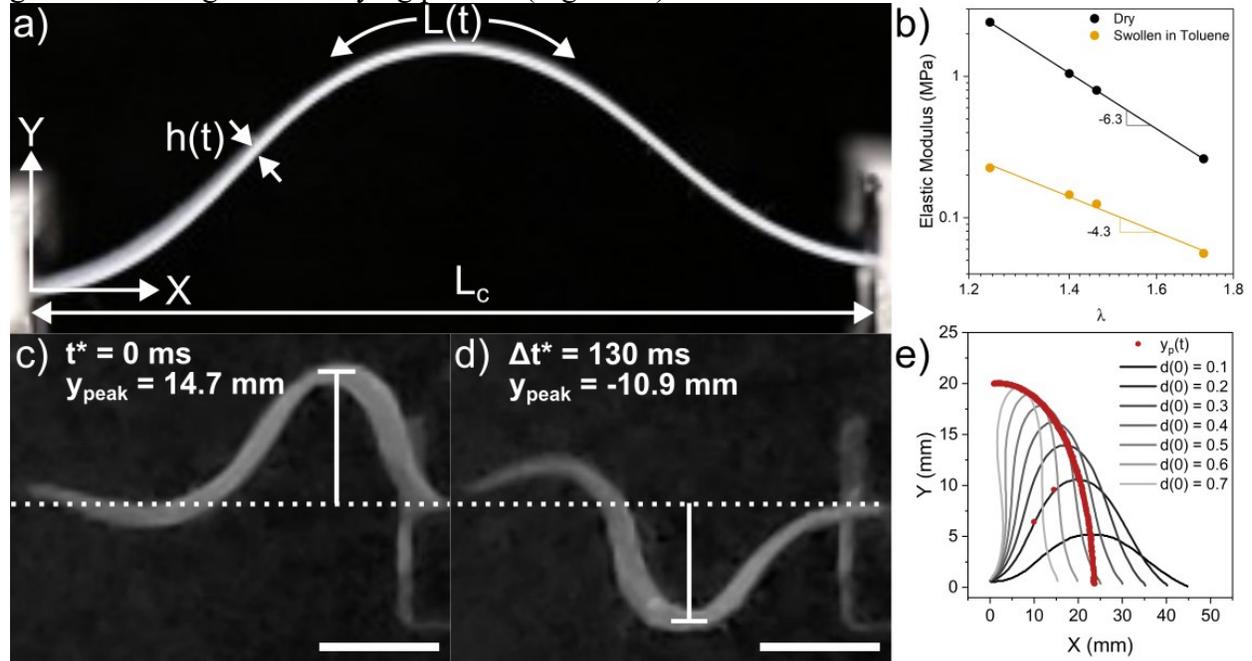

**Figure 2.** The profile of the swollen beam is dependent on the initial confinement and elastic properties of the crosslinked network. (a) An annotated image of a buckled beam illustrating the time-dependent geometric properties and confined length ($L_c$) of the swollen actuator. (b) The dry and swollen elastic moduli, measured using indentation, are plotted with the experimental linear equilibrium swelling ratio ($\lambda$). The profile of the beam during snapping is illustrated as a (c) pre- and (d) post-snap profile with annotated peak displacements at a relative snap duration, Δt*,

and (e) the extracted beam profile for the $E(\infty) = 0.3$ MPa network at the seven tested initial confinements ($d(0)$). The peak point ($y_p$) along the profile is plotted as a function of increasing initial confinement.

The dry confinement ($d(\infty)$) can be calculated similarly to the initial confinement as the difference in the dry length of the beam and the confined length, normalized by the dry length of the beam (Equation 1b).

$$d(\infty) = \frac{L(\infty) - L_c}{L(\infty)} \qquad \text{Equation 1b}$$

Combining Equation 1a & 1b by substitution through the confined length produces a relationship between the swelling property of the crosslinked network and the geometric confinement of the experiment shown in Equation 1c.

$$\lambda = \left[\frac{1 - d(\infty)}{1 - d(0)}\right] \qquad \text{Equation 1c}$$

We use the geometric properties of the confined beam in addition to the polymer network properties to characterize the performance of the autonomous snapping beam. We analyze this performance on two criteria: snapping repeatability and power. Both criteria are measured from an equilibrium swollen state until the beam is completely dried.

**Snap repeatability**

A swollen polymer gel is capable of autonomously snapping when confined because the convex surfaces, which are under tension, have relatively greater surface evaporation than the concave counterpart.[17] The beam is capable of multiple consecutive snaps due to the reversibility of the buckled strain gradient, switching the direction of the diffusive driving force. We characterize the snap repeatability through the snap lifetime and snap count, which describe the time until the final snap and the total number of snaps, respectively, for a fully swollen beam to completely dry.

The snap lifetime defines the amount of time between the beginning of the experiment and the final snap. The snap lifetime is expected to be a function of the time required to fully dry the beam and the energy required to generate a snap at low solvent concentrations. The total solvent loss as a function of time is a poroelastic process that is independent of the initial lateral confinement. We measure the solvent mass fraction, $M_s$, during a variety of experimental conditions. The solvent mass fraction is plotted as a function of time, normalized by the poroelastic timescale of the beam, $\tau = \frac{(\lambda h(\infty))^2}{4D}$, which identifies a common function for solvent mass fraction for all experimental conditions (Figure 3a). The poroelastic timescale uses the equilibrium swelling ratio to account for differences in size and solvent mass between the different crosslinked networks. A snap lifetime of zero means that no snaps occurred during the entire drying process. We use snap lifetime to gauge the duration of time where solvent evaporating from the swollen beam can produce an autonomous snap. The longest snap lifetime, 852 s, occurred for $E(\infty) = 0.8$ MPa and $d(0) = 0.6$, with a total of five snaps over the period of drying. The results for all 28 experimental conditions are summarized in Figure 3b and 3c. The snap lifetime increases as the crosslink density decreases. Networks that are less densely crosslinked are capable of amassing more solvent when fully swollen.[18] Consequently, the time to evaporate all solvent increases as crosslink density decreases, suggesting that decreasing crosslink density will also increase snap lifetime. Yet some experimental conditions, such as $E(\infty) = 0.8$ MPa and $d(0) = 0.5$, did not snap at long times, resulting in a much lower snap lifetime. From this we conclude that the snap lifetime is not only a performance of the availability of solvent, but also the capability for the residual evaporating solvent to overcome a transient snap-through energy barrier.

We compare the duration of the first snap and all snaps in each experiment recorded using the highspeed camera in Figure 3d. The snapping duration for all snaps across 28 experiments had an average of $0.13 \pm 0.05$ s. Notably, the first snap in each experiment, which was the most powerful in 23 of the 24 experiments that produced a snap, similarly had an average snapping time of $0.12 \pm 0.03$ s. These findings demonstrate that the timescale for the duration of a snap is consistently around 0.12–0.13 s despite the first snap initiating from a homogenous, swollen state whereas consecutive snaps begin from a post-snap concentration gradient. This measured snapping timescale interestingly matches the transversal flow timescale that controls the snapping closure in Venus flytraps.[8] While the snap lifetime is useful in characterizing the capability for an autonomous snap, the total number of snaps describes the proficiency a given experimental condition is able to produce autonomous snaps.

The snap count defines the total number of snaps. Within the range of crosslinked networks and initial confinements tested, $E(\infty) = 0.8$ MPa, $d(0) = 0.3$ had the greatest snap count, completing 49 consecutive autonomous snaps. From the recording of the beam profile captured at 30 fps, we use a custom MATLAB script to extract the beam profile as solvent evaporates (Figure 4a). The drying beam is initially symmetric, but as solvent evaporates and the beam undergoes snapping, the concentration profile becomes complex, producing asymmetric, higher order buckling modes. While this behavior is unique to these evaporating structures, describing this evolving profile requires rigorous modelling of the solvent concentration profile that is outside the scope of the present results. The snap count for each condition is plotted in Figure 3e & 3f, identifying an optimal region for maximizing the snap count when varying the dry modulus and initial confinement. All conditions snapped at least once except for $E(\infty) = 2.4$ MPa for $d(0) > 0.3$, where the stiffness of the buckled beam was greater than the evaporative driving force to overcome the energy barrier to snap.

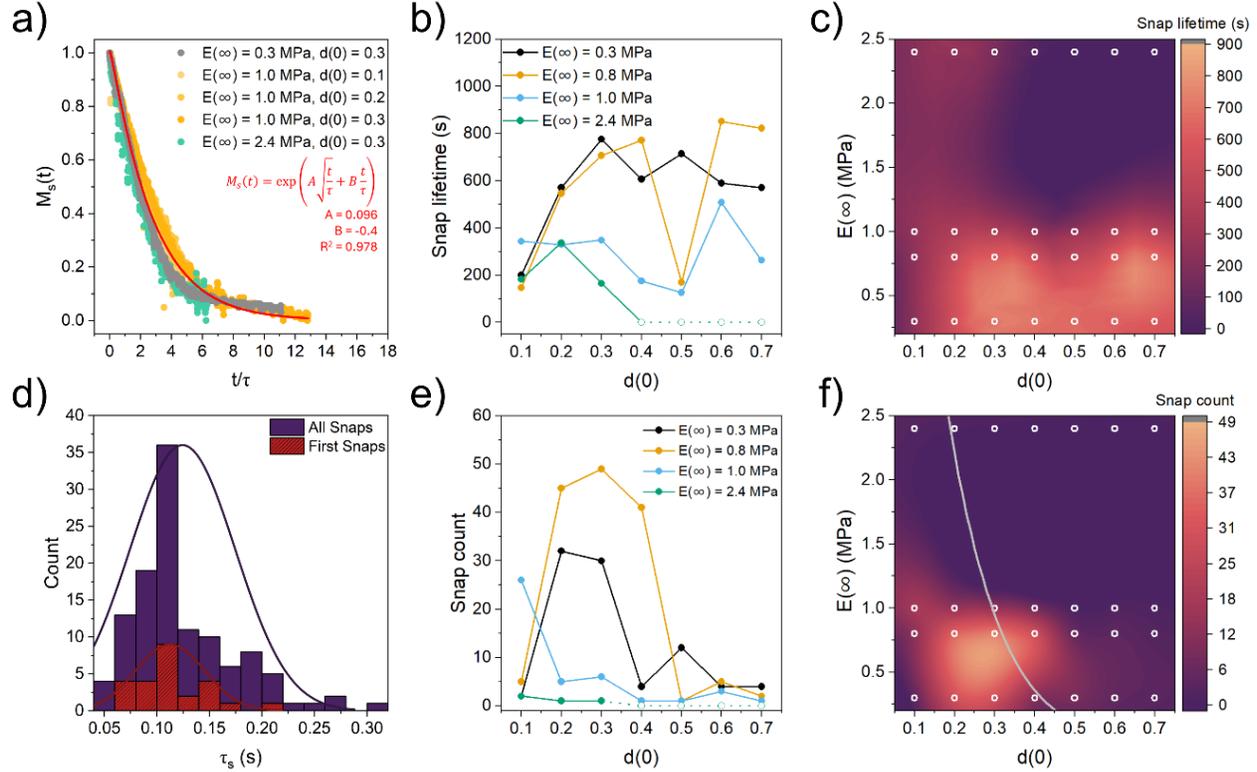

**Figure 3.** (a) Normalized solvent mass fraction loss ($M(t)$) over time ($t$) with a consolidated fit of five mass loss experiments. The poroelastic time ($\tau$) increases as the equilibrium swelling ($\lambda$) increases, so this normalization consolidates the differences in equilibrium solvent mass and volume. Snap lifetime as a (b) scatter plot and (c) contour plot as a function of dry modulus ($E(\infty)$) and initial confinement ($d(0)$) identify that the snap lifetime is proportional to the poroelastic time. (d) Distribution of the snap duration ($\tau_s$) across all snaps and the first snaps for each experimental condition. Snap count as a (e) scatter plot and (f) contour plot as a function of dry modulus and initial confinement identify that the greatest snap count values are localized to specific experimental conditions. The zero-strain condition (solid gray line) is overlaid to illustrate the intersection of this optimization.

We found that the time between consecutive snaps decreased by an order of magnitude during the snap lifetime, suggesting that the driving force to snap is less dependent on the amount of solvent that evaporates and more dependent on the development of a concentration gradient. During drying, the beam is shortening and stiffening, both of which affect the energy barrier to snap, but because the time between snaps decreases over time, we expect the length-shortening, or decreasing confinement, to dominate in promoting snapping. The length of the drying beam over time for the E(∞) = 0.3 MPa, d(0) = 0.2 condition is plotted in Figure 4c. These results show that initial confinement and swelling are critical for controlling the lifetime and count of the autonomous snapping system.

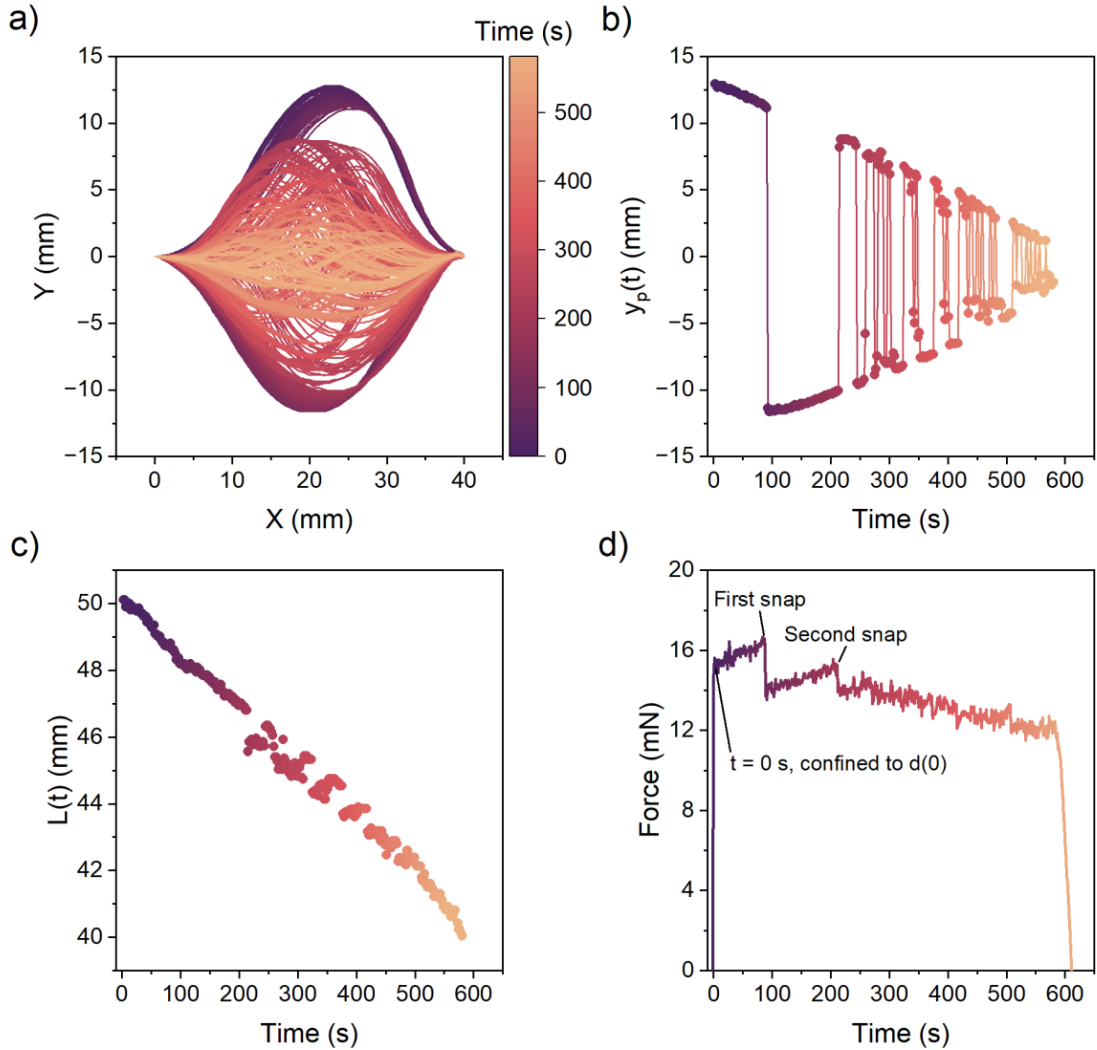

**Figure 4.** (a) The profile of the drying beam extracted using a custom MATLAB script from the 30-fps video as solvent evaporates, each line plotted represents the profile captured at 1 s intervals. The profile is characterized by the (b) peak point ($y_p(t)$) along the profile over time and the (c) swollen beam length ($L(t)$) as a function of time. (d) The lateral force response ($F(t)$) at the grips as a function of time is used to calculate the power of individual snaps.

We have shown that the snap lifetime and count are a function of the material and geometric properties of the swollen, buckled beam. Additionally, we identify that the driving force for snapping is dependent on poroelastic properties and the confinement state as the beam dries; as a result, the snapping lifetime is primarily dictated by the crosslink density of the network, while the snap count is a continuous balance of the confinement and the snap-through energy barrier. Furthermore, the capability for a drying beam to repeatedly snap is improved by increasing the available solvent mass for evaporation and by appropriately confining the beam to maximize how much the evaporating solvent contributes directly to snapping.

**Snap power**

The snap power quantifies the rate of energy released during snapping. This energy accumulates as the solvent evaporates and produces an asymmetric concentration profile that induces a bending energy on the buckled structure as the loading time increases (~$10^1$–$10^2$ s, dependent upon modulus, confinement, and the duration of the experiment). Once this bending energy overcomes the energy barrier to snap, the swollen beam releases this energy in shorter time intervals (~$10^{-1}$ s) than the loading period. Maximizing snap power becomes a balance of profile stability, or the resistance to perturbation, and evaporative asymmetry. The snap power ($P$) is defined in Equation 2 as the product of the change in lateral force response ($\Delta F$) and the absolute peak displacement before and after a snap ($|\Delta y_p(t)|$) divided by the snap duration ($\tau_s$).

$$P = \frac{\Delta F(t) \cdot |\Delta y_p(t)|}{\tau_s} \qquad \text{Equation 2}$$

The snap duration is independent of experimental condition and the time at which the snap occurs (Figure 3d); the average snap duration for both the first snap and all snaps for a given experiment is about 0.12 s. Consequently, the power of an individual snap is primarily a product of the change in lateral force and peak displacement. While both the lateral force and peak displacement are dependent on the transient modulus and lateral confinement, we plot the maximum power from each experimental condition as a scatterplot in Figure 5a and as a contour plot in Figure 5b. The maximum power increases with increasing initial confinement. The most powerful snap was 4.61 mW at $E(\infty)$ = 1.0 MPa, $d(0)$ = 0.7. The maximum power data is interpolated in a contour plot in Figure 5b to show the maximum power is predicted to be between $E(\infty)$ = 1.0 MPa and $E(\infty)$ = 2.4 MPa for $d(0) > 0.7$. Like the snap count, snap power can be optimized as a function of the initial confinement and dry modulus of the beam. Furthermore, we have shown that snap power is independent of how efficiently solvent evaporation contributes to snapping, but rather is a function of the maximum confinement, or snap energy barrier, that the evaporating solvent is capable of overcoming.

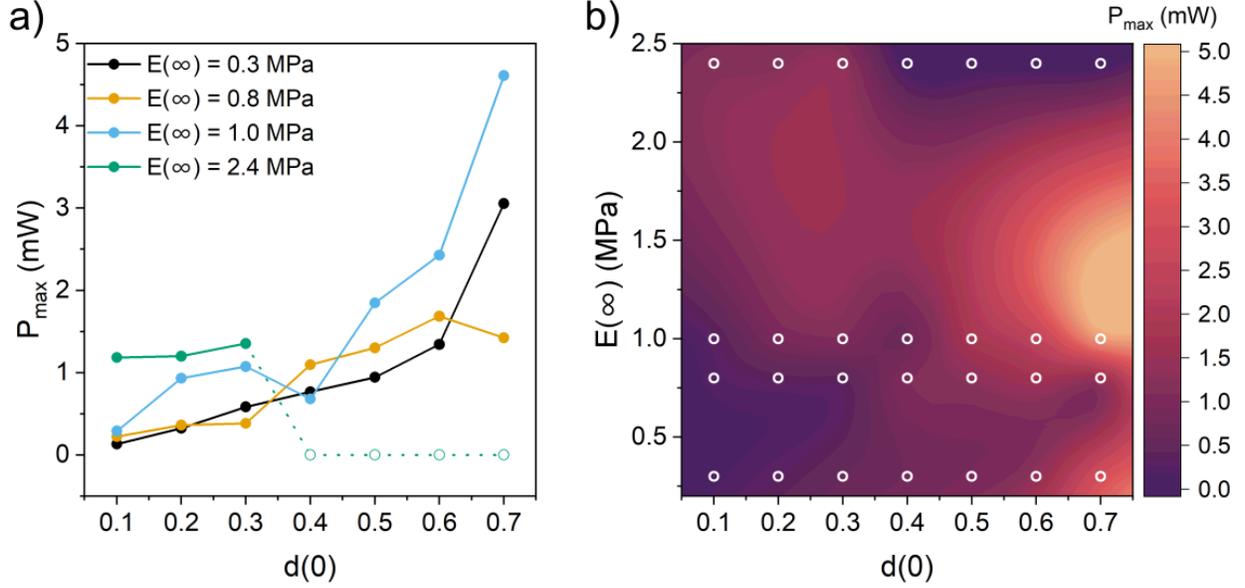

**Figure 5.** Maximum snap power ($P_{max}$) for all experimental conditions as a (a) scatter plot and (b) contour plot.

**Optimizing snap repeatability and power**

Snap repeatability and power depend upon material properties and geometric confinement. We present two optimization principles that guide the design of autonomous snapping gel through solvent efficiency and a cumulative, weighted snap resistance. The dry modulus of the PDMS network can be related to the equilibrium swelling through Flory-Rehner theory, where the equilibrium swelling is a balance of osmotic and elastic free energy.[18,19] Accordingly, substituting Equation 1c into the Flory-Rehner relation of dry modulus and equilibrium swelling, the dry modulus can be written as a function of the Boltzmann constant ($k$), temperature ($T$), Kuhn length ($b$), and geometric confinement of the beam (Equation 3a).

$$E(\infty) \cong \frac{kT}{b^3}\left[\frac{1-d(0)}{1-d(\infty)}\right]^5 \qquad \text{Equation 3a}$$

As solvent evaporates, the total length of the beam decreases while the length between the grips is constant, so the transient confinement of the drying, buckling beam decreases monotonically with solvent concentration. Because solvent transport is the driving force for snapping, for our first optimization principle, we propose a solvent efficiency condition where the dry beam is under zero strain at long times, ($d(\infty) \to 0$), allowing for the maximum amount of solvent evaporation, or input energy, to contribute to snapping as the beam approaches a minimum energy state. This "zero-strain" condition ($d(0)^*$) is an initial confinement that is a function of the linear equilibrium swelling ratio, shown in Equation 3b:

$$E(\infty) \cong A \frac{kT}{b^3}[1 - d(0)^*]^5 \qquad \text{Equation 3b}$$

We add a fitting parameter ($A$) to fit the equation to our calculated values of $d(0)^*$ using the experimentally measured linear equilibrium swelling ratios for each crosslinked network (Figure S1). Consequently, if $d(0) > d(0)^*$, the resulting dry beam has a length greater than the distance between the grips and remains buckled. As a result, a greater bending energy remains in the buckled beam that resists snapping more than the unbuckled state. Similarly, if $d(0) < d(0)^*$, the drying beam, at a certain point in time, will approach a state of tension while partially swollen. The remaining solvent in the partially swollen beam does not contribute to snapping and so the capacity for the drying beam to snap is prematurely terminated. We plot Equation 3b on Figure 3f and find good agreement for $E(\infty) \leq 1.0\ MPa$ with where the greatest snap counts intersect with the zero-strain condition. The zero-strain condition guides the selection of an optimal geometric confinement for a given linear equilibrium swelling ratio but does not consider any properties of snap-through buckling.

For creating the second optimization principle, we use the results for the normalized critical snap-through force as a function of lateral confinement, $\bar{F}(d(t))$, from Zhang et al. and our normalized solvent mass-loss data to calculate a weighted arithmetic mean that spans the transient confinement as $d(0) \to d(\infty)$:

$$\Pi = \frac{\int_{d(0)}^{d(\infty)} \bar{F}(d(t)) \cdot M_s(d(t)) dd(t)}{\max(d(0), d(0) - d(\infty)) \int_{d(0)}^{d(\infty)} M_s(d(t))\, dd(t)} \qquad \text{Equation 4}$$

The effective critical snapping force, $\Pi$, represents the cumulative resistance to snapping for a given $E(\infty)$ and $d(0)$. We fit our fractional solvent mass-loss data from Figure 3a to a general mass-loss equation concerning diffusion and evaporation to obtain a normalized function, $M_s(d(t))$, that represents the fractional solvent concentration over a normalized time using the poroelastic time of the beam ($\tau$). This derivation is derived in more detail in supplementary information and Figure S2.

Describing the concentration profile over the entire snapping process is complex, so we assume the evolution of solvent-dependent properties (namely, $E(t), d(t), w(t), L(t)$, and $h(t)$) is proportional to $M_s(d(t))$. This normalization function is used as the weighting for $\bar{F}(d(t))$ because of the guiding principle that solvent transport is the key driving force for autonomous snapping. The range for $\Pi$ is calculated by $\max(d(0), d(0) - d(\infty))$ and is representative of the zero-strain condition such that the resistance to snapping is minimized by maximizing the amount of solvent evaporation that occurs when the beam is buckled. We present a contour plot of $\Pi$ in Figure 6 with annotations describing how cumulative resistance to snapping changes as a function of swelling, elasticity, and confinement. As an experimental condition approaches the zero-strain condition, the cumulative snap resistance decreases. Coupling this with our results for the snapping lifetime provides an explanation for the localization of the optimal snap count shown in Figure 3f.

As crosslink density decreases, the linear equilibrium swelling ratio increases, increasing both the poroelastic timescale and the solvent mass within the fully swollen beam. Even though the cumulative resistance to snap decreases as $E(\infty)$ increases, the snap count increases as more solvent is available to drive snapping at lower values of $E(\infty)$. The optimization for snap count is thus defined by the balance between an increasing cumulative resistance to snapping with an increasing availability of solvent.

Our results demonstrate that snap power increases with greater initial confinement, reflecting the direct scaling of snap power with peak beam displacement. While the peak of a buckled profile plateaus with increasing confinement, increasing confinement continues to increase the maximum snap power so long as the solvent-induced evaporative stress is sufficient to initiate snap-through for a given beam stiffness. The maximum snap power increases with greater cumulative snap resistance as a greater applied force will be accumulated prior to snapping. To extend these lessons to other material systems, the relationship between the evaporation stresses and an effective applied load on the beam would need to be further investigated. Here we assume the distributed evaporative stress along the length of the beam acts as a point force at the midpoint of the beam, consistent with the referenced literature.

**Discussion**

The autonomous snap-through performance of swollen polymer beams has been characterized as a function of dry modulus and lateral confinement. The performance of the snapping beams is defined by the snap repeatability, through the snap lifetime and count, and the maximum snap power. We found that the snap lifetime relies upon the poroelastic time of the solvent swollen beam, whereas the snap count is an optimization of the poroelastic time and the effective critical snapping force. We design guiding principles for maximizing the performance of autonomous snapping beams based on their capacity to repeatably actuate and the maximum power of an individual actuation. Because a swollen autonomous actuator has a finite solvent mass for generating snaps, a resource allocation trade-off determines whether the drying beam will generate a few powerful snaps or many weak snaps. We show how the poroelastic time and effective critical snapping force guide the design of autonomous actuation in maximizing the desired performance.

While we expect these guiding principles to apply to other geometries, such as swollen shells, or other polymer-solvent material systems, these effects on performance have yet to be determined. The polymer network of the swollen beam follows a typical swelling response, where the elastic modulus is inversely related to solvent volume fraction. Other molecular polymer architectures like bottlebrush networks have been found to stiffen with increasing solvent concentration, which may provide improved power and repeatability performance for better tuning transient properties to geometric boundary conditions. Describing the differences in these systems will require a detailed understanding of the evolution of solvent concentration gradient in the snapping beam. Modelling of the transient concentration gradient could provide insight into how the evaporating beam reaches an instability prior to snapping. While we have shown that solvent transport is crucial to actuation performance, a question remains on whether solvent evaporation along the length of the beam equally contributes to snapping. Characterization of a heterogeneous beam profile and transient material properties could improve the efficiency of solvent transport and actuator performance and extend the material system to more solvents, like water, that are desirable for environmental applications and renewability.

**Conclusion**

We have mapped the snapping metrics of snap count, lifetime, and power to capture the repeatability and power-amplification of evaporation-driven snapping elastomers by connecting the autonomous actuation mechanism to the properties of the polymer network and geometric confinement. Continuous snap-through actuation is improved by optimizing solvent evaporation with the zero-strain condition and by maximizing the snap lifetime relative to the effective critical snapping force. This condition is expected to be general, extending to geometries beyond beams, including shells, which can be used for motions that include jumping.[20,21] These guidelines will help to design energy-efficient, power dense, devices, including robotic systems and self-regenerating seed dispersal systems. The results reveal a rich landscape of materials structures and macroscale constraints to be explored to synergistically design autonomous motion that pushes the limits of performance.

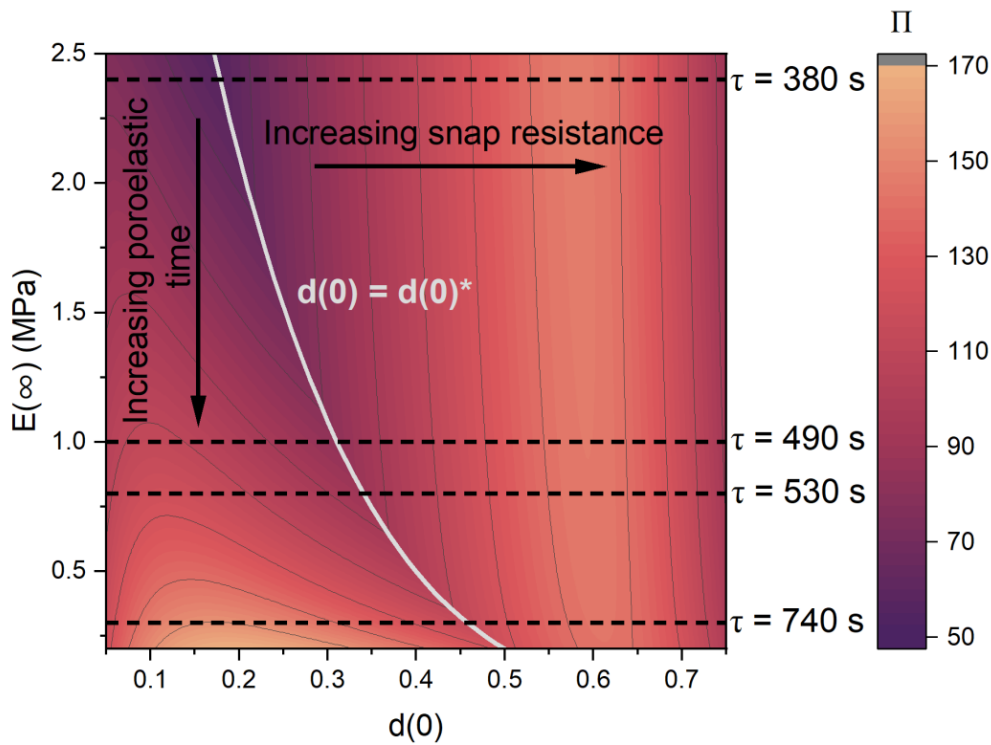

**Figure 6.** Annotated contour plot of the effective critical snapping force ($\Pi$) calculated from a weighted arithmetic mean of the normalized critical snapping force ($\bar{F}(d(t))$) over the range of transient confinements in the drying beam and the distribution of solvent mass as a function of time. The effective critical snapping force decreases as the experimental conditions near the zero-strain condition ($d(0)^*$) where solvent transport is optimized. The poroelastic time ($\tau$) for each crosslinked network is annotated on the righthand side of the contour plot.

## Materials and Methods

**Polydimethylsiloxane (PDMS) ribbon.** PDMS ribbons were made using a commercial silicone base and crosslinker (Sylgard 184). Base to crosslinker ratios of 30:1, 20:1. 10:1, and 5:1 were mixed, degassed, and cured at 70 °C for at least 24 hours in a 3D printed PLA mold. Mold thickness was used to control thickness of the PDMS sheet to 0.5 mm. Individual ribbons were laser cut from the mold in dry dimensions of 5mm x 50 mm and the free chains were extracted over five swelling and drying cycles in toluene. Multiple ribbons were made from single sheets, which were used across multiple experiments. Four crosslinking ratios of polydimethylsiloxane (PDMS, Sylgard 184) were molded into 0.5 mm thick sheets. A laser cutter was used to extract rectangular ribbons with a width and length of 5 mm and 70 mm, respectively. Each ribbon was swollen and dried three times in toluene to extract uncrosslinked polymers.

**Solvent mass fraction loss over time.** A custom 3D printed stage was designed to clamp swollen ribbons on both ends at a fixed distance used spring grips to ensure an active grip with a deswelling thickness. The stage was placed on a mass balance (OHAUS Explorer Series Precision Balance, capacity=1.1kg, Readability = 0.001g) such that the evaporating mass of the ribbon was measured under similar testing conditions as the tensile tests on the Texture Analyzer.

**Force Response and Capturing Beam Profile.** To determine the reactionary force of the confined, swollen beam, each sample was confined in clamped grips with an initial separation of 50mm on a Texture Analyzer (Stable Microsystems). The grips confined the beam at 10mm/s until the experimental confinement was reached and reactionary force response was measured over time. While the Texture Analyzer measured the force response of the evaporating beam, a camera captured the entire deswelling process at 30 frames per second while a highspeed camera was repeatably engaged to capture any high-speed motion at 960 frames per second. Both videos were rendered through Blender to extract individual frames of the experiment during which they were cropped and compressed for ease in data handling. ImageJ was used to auto determine thresholding parameters for each experiment and measure the image scale. A custom MATLAB script was written to extract the profile and geometric properties of the beam over time. A smoothing function was used to determine geometric curvature of the beam and identify when snapping occurs.

## Acknowledgments
This work is supported by the U. S. Army Research Laboratory (W911NF-15-1-0358), the U. S. Army Research Office (W911NF-23-2-0022), and NSF CAREER Award (No. 1845631).

## Supplementary Information

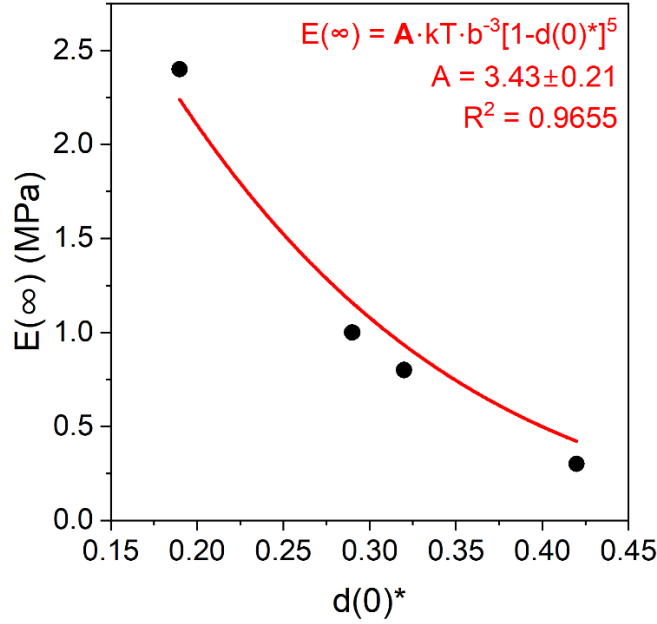

**Figure S1.** Fitting of Equation 3b to the experimental data for dry modulus ($E(\infty)$) and linear equilibrium swelling ratio ($\lambda$). The zero-strain condition ($d(0)^*$) is calculated from Equation 1c as the long-time confinement approaches zero ($d(\infty) \to 0$) using the experimentally measured linear equilibrium swelling values shown in Figure 2b.

**Derivation of the effective critical snapping force (Π).**

We assume that the transient properties of the swollen, evaporating beam are locally independent and focus on the overall change in confinement and mass loss over time with respect to the normalized critical snapping force characterized by Zhang et al.[22] This normalized critical snapping force was a result of a point load at the midpoint of the beam, which changes as a function of the lateral confinement of the beam. As the swollen beam evaporates solvent, the beam length over time ($L(t)$) decreases such that the lateral confinement ($d(t)$) also decreases as a function of time.

We calculate a weighted arithmetic mean of the normalized critical snapping force versus confinement using the distribution of mass loss over a normalized time. At $t = 0$, $d(t) = d(0)$ and $M_s = 1$. As $t \to \infty$, $d(t) = d(\infty)$ and $M_s = 0$.

The average normalized critical force is calculated in Equation S1.

$$\bar{F}_{avg} = \frac{\int_{d(0)}^{d(\infty)} \bar{F}(d(t)) \, dd(t)}{\max\left(d(0), d(0) - d(\infty)\right)} \qquad \text{Equation S1}$$

$\bar{F}_{avg}$ is a simple mean of the values for the normalized critical snapping force and the range over which the beam is confined. This range is defined by $\max\left(d(0), d(0) - d(\infty)\right)$ and is a description of the zero-strain condition such that if $d(0) > d(0)^*$ the dried beam profile remains buckled, $d(\infty) > 0$, and the range which the normalized critical snapping force should be averaged is $d(0) - d(\infty)$. If $d(0) < d(0)^*$, the dried beam profile is under tension, $d(\infty) < 0$, and the range which the drying beam was capable of snapping would be $d(0)$. However, this function does not account for how the evaporating mass is distributed over time throughout the experiment. Most solvent evaporates at shorter time scales due to a larger concentration gradient with the ambient environment.

To account for the time-dependent mass distribution, we use the normalized mass-loss data as weighting for $\bar{F}_{avg}$. The function for the solvent mass fraction is linearly projected to the transient lateral confinement such that the range of lateral confinement as $d(0) \to d(\infty)$ is proportional to $\left[M_s\left(\frac{t}{\tau}=0\right)=1\right] \to \left[M_s\left(\frac{t}{\tau}=\infty\right)=0\right]$.

Adding the mass-loss weighting produces Equation 4 (Equation S2):

$$\Pi = \frac{\int_{d(0)}^{d(\infty)} \bar{F}(d(t)) \cdot M_s(d(t)) dd(t)}{\max(d(0), d(0)-d(\infty)) \int_{d(0)}^{d(\infty)} M_s(d(t)) dd(t)} \qquad \text{Equation S2}$$

Each experimental condition calculates an average normalized critical snapping force that is weighted by the transient amount of mass loss during drying to describe an effective critical snapping force from the initial confinement and equilibrium swelling properties of the network.

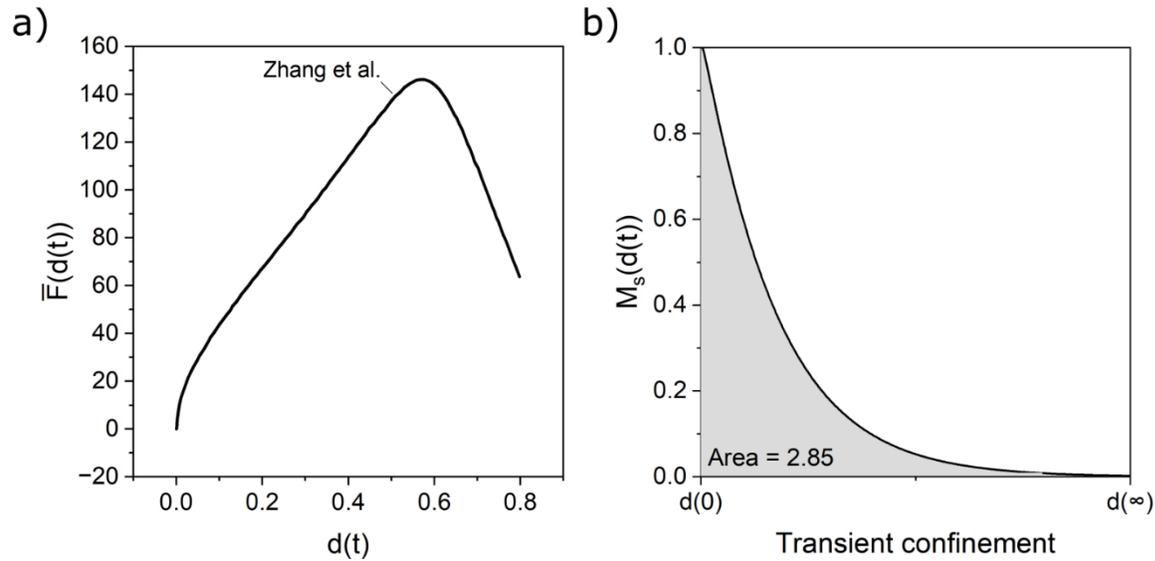

**Figure S2:** (a) normalized critical snapping force ($\bar{F}(d(t))$) as a function of lateral confinement ($d(t)$). Here we show the lateral confinement as a function of time to account for the transient confinement as the normalized critical snapping force decreases as the beam length dries and the confinement decreases. (b) The normalized fractional mass loss data as a function of time is linearly projected to the change in lateral confinement, which is used as a weighted for the normalized critical snapping force to normalize the distribution of solvent loss across the entire experiment.